# LaBaNiO$_4$: A Fermi glass

**A Schilling[1,5], R Dell'Amore[1], J Karpinski[2], Z. Bukowski[2], M Medarde[3,4], E Pomjakushina[3,4] and K A Müller[1]**

[1] Physics Institute of the Universität of Zürich, Winterthurerstrasse 190, CH-8057 Zürich
[2] Laboratory for Solid-State Physics, ETH Zürich HPF F-7, CH-8093 Zürich
[3] Laboratory for Neutron Scattering, ETHZ & PSI, CH-5232 Villigen PSI
[4] Laboratory for Developments and Methods, ETHZ & PSI, CH-5232 Villigen PSI

Email: schilling@physik.uzh.ch

**Abstract.** Polycrystalline samples of LaSr$_{1-x}$Ba$_x$NiO$_4$ show a crossover from a state with metallic transport properties for $x = 0$ to an insulating state as $x \to 1$. The end member LaBaNiO$_4$ with a nominal nickel Ni $3d^7$ configuration might therefore be regarded as a candidate for an antiferromagnetic insulator. However, we do not observe any magnetic ordering in LaBaNiO$_4$ down to 1.5 K, and despite its insulating transport properties several other physical properties of LaBaNiO$_4$ resemble those of metallic LaSrNiO$_4$. Based on an analysis of electrical and thermal-conductivity data as well as magnetic-susceptibility and low-temperature specific-heat measurements we suggest that LaBaNiO$_4$ is a Fermi glass with a finite electron density of states at the Fermi level but these states are localized.

---

[5] Corresponding author

# 1. Introduction

Certain transition-metal oxides show unique physical properties that are not only of academic interest but also make them technologically useful. The discoveries of high-temperature superconductivity [1] and colossal magnetoresistance [2] in such oxides have triggered a tremendous activity in the research into these compounds with the aim to understand their peculiar magnetic and transport properties.

The metallic state in the high-temperature superconductors, for example, is highly unusual because it evolves from originally electrically insulating oxides. Stoichiometric $La_2CuO_4$ is antiferromagnetic near room temperature, with copper in a Cu $3d^9$ (spin $S = 1/2$) configuration, and it is regarded as a Mott insulator. These features are generally believed to be important ingredients for the occurrence of superconductivity on subsequent doping the compound with hole-type charge carriers, e.g., by the partial substitution of La by Ba or Sr [1]. By contrast, stoichiometric $LaSrNiO_4$, that shares the same $K_2NiF_4$ structure with $La_2CuO_4$, is metallic [3], although a $3d^7$ configuration of nickel in a low-spin $S = 1/2$ state might be expected that could also lead to an antiferromagnetic and insulating state. This fact has been explained with the partial occupation of the oxygen O $2p$ energy band of $LaSrNiO_4$ by hole states that hybridize with the nickel Ni $3d^8$ states, in very contrast to $La_2CuO_4$ that shows a completely filled copper Cu $3d^9$ lower-Hubbard band [4]. The occupation of oxygen $2p$ hole states, leaving the nickel in a Ni $3d^8$ configuration, is very common in nominally trivalent-nickel oxides [5,6,7]. For this reason $LaSrNiO_4$ has not been considered to be a serious candidate for a "parent compound" of superconductors, and substitution experiments in $La_{2-y}Sr_yNiO_4$ over a wide range of doping (i.e., 0 $\leq y \leq$ 1.5) have not resulted in superconducting samples [3,8].

In very contrast to the metallic character of $LaSrNiO_4$ the isostructural compound $LaBaNiO_4$ has been reported to be an insulator at zero temperature $T$, and to undergo a transition from a high-spin to a low-spin $S = 1/2$ state of Ni around $T \approx 120$ K with decreasing temperature [9]. It is therefore of particular interest to study the influence of a partial or a complete substitution of Sr by Ba in $LaSrNiO_4$ on the physical properties that are expected to be affected by a resulting stretching of the crystal lattice [9,10]. We have found that there exists a solid solution of Ba in $LaSr_{1-x}Ba_xNiO_4$ for $0 \leq x \leq 1$, and we have measured the electrical and the thermal conductivities, the magnetic susceptibilities and the low-temperature specific heats of polycrystalline $LaSr_{1-x}Ba_xNiO_4$ samples. Since possible magnetic-ordering phenomena as well as other physical properties may strongly depend on the exact oxygen stoichiometry, we have combined a thorough neutron diffraction analysis of the end member LaBaNiO$_4$ with a chemical analysis and with additional heat-treatment experiments at an elevated oxygen pressure.

## 2. Experimental

We have prepared polycrystalline samples of LaSr$_{1-x}$Ba$_x$NiO$_4$ for $x = 0, 0.25, 0.3, 0.35, 0.4, 0.45, 0.5, 0.75$ and $1.0$ by a standard wet chemical procedure. Mixtures of corresponding metal nitrates were dissolved in nitric acid. The liquid was then slowly evaporated, and the remaining mixture of nitrate powders was pre-reacted at 900 °C for several hours in air. The remaining mixture of oxides was pressed into pellets and sintered at 1100 °C during 3 days in air. Samples of LaSr$_{0.5}$Ba$_{0.5}$NiO$_{4-\delta}$ ($x = 0.5$) and LaBaNiO$_{4-\delta}$ ($x = 1.0$) were post-annealed at temperatures between 400 °C and 600 °C and at an O$_2$ pressure of 500–880 bar (50–88 MPa). After testing the phase purity of all the samples with a Guinier camera using Cu-$K_{\alpha 1}$ radiation we have performed a thermogravimetric analysis on 100 mg of each as-prepared and high-pressure oxygen annealed LaBaNiO$_{4-\delta}$, followed by neutron-diffraction experiments on 5 g powdered samples of the same batches at various temperatures between 1.5 and 550K. These experiments were performed at the neutron source SINQ in Villigen, Switzerland, on the powder diffractometers HRPT ($\lambda = 0.15$ and $0.18$ nm) and DMC ($\lambda = 0.245$ and $0.4$ nm). The four-probe electrical conductivities and the AC magnetic susceptibilities were measured for all the samples in zero DC magnetic field. Specific-heat data were taken on three samples ($x = 0$ and high-pressure oxygen annealed $x = 0.5$ and $x = 1$), while the thermal conductivities were measured only on the $x = 0$ and the high-pressure oxygen annealed $x = 1$ samples. All the physical properties were collected using standard factory options in a commercial PPMS platform (Physical Property Measurement System, Quantum Design Inc., San Diego, USA).

## 3. Results

*3.1. Chemical analysis and neutron-diffraction experiments*

A thermogravimetric analysis of the as-prepared $x = 1$ sample LaBaNiO$_{4-\delta}$ yielded an oxygen content 3.85(1) ($\delta \approx 0.15$) assuming a nominal cation ratio La:Ba:Ni = 1:1:1. Under the same assumption, a Rietveld refinement of the room-temperature neutron-diffraction pattern using the space group I4/mmm proposed in previous work [9,10] (see figure 1) also gave an oxygen content 3.85(1), with oxygen vacancies located both in the apical positions ($\approx 33\%$) and in the basal planes ($\approx 66\%$). A refinement with the restriction (La+Ba):Ni = 2:1 gave similar $R$-factors and suggested a composition La$_{1.27(2)}$Ba$_{0.73(2)}$NiO$_{3.97(2)}$ which would imply a considerable amount of unreacted BaO present as an impurity. However, we could not find any detectable amount of BaO and/or its carbonated subproducts, neither in the diffraction patterns nor by chemical methods. To clarify this issue we repeated the chemical analysis and the neutron-diffraction measurements on samples that were annealed at 400 °C under an oxygen pressure of 880 bar (88 MPa) for 50 hours (annealing at 600 °C at elevated oxygen pressure resulted in a partial decomposition of the sample). According to the observed weight change we estimated an oxygen uptake of $\approx$ 0.16/formula unit during this procedure, which is consistent with the above value $\delta \approx 0.15$ for the as prepared sample. The thermogravimetric analysis of the high-pressure oxygen-annealed sample resulted in an oxygen content 4.025(10), thereby confirming the nearly stoichiometric composition $LaBaNiO_4$ after the high-oxygen pressure treatment. The structural parameters of oxygen-annealed $LaBaNiO_4$ as obtained from DMC data refinements ($\lambda = 0.245$ nm) using the tetragonal space group I4/mmm are listed in table 1 for three different temperatures $T = 1.5$ K, 295 K and 550 K, respectively.

**Table 1.** Lattice parameters and refined atomic coordinates of $LaBaNiO_4$ using the space group I4/mmm. (La,Ba): 4e (0 0 $z_1$); Ni: 2a (0 0 0); O1: 4e (0 0 $z_2$); O2: 4c (0.5 0 0).

|         | $T = 1.5$ K     | 295 K           | 550 K           |
|---------|-----------------|-----------------|-----------------|
| $a$ (nm) | 0.384970(11)   | 0.385519(11)    | 0.386754(11)    |
| $c$ (nm) | 1.273670(56)   | 1.278893(57)    | 1.286938(58)    |
| $z_1$   | 0.36086(62)     | 0.36050(63)     | 0.36080(64)     |
| $z_2$   | 0.16497(57)     | 0.16509(59)     | 0.16568(58)     |

As expected and also reported in [9,10], the lattice parameters of $LaBaNiO_4$ are larger than those of $LaSrNiO_4$ ($a = 0.3826$ nm, $c = 12.45$ nm [11]), and vary smoothly from $x = 0$ to $x = 1$ according to our Guinier Cu-$K_{\alpha 1}$ diffraction patterns. In agreement with a report by Alonso *et al.* [10] we detected a distinct extra reflection peak in the neutron-diffraction data of the as-prepared $LaBaNiO_{4-\delta}$ sample that can be indexed as (1/2, 0, 2), indicating a doubling of the unit cell in the *a*-direction and changing the crystal symmetry from tetragonal to orthorhombic. Upon high-pressure oxygen annealing this peak slightly moved to an incommensurable position (0.516, 0, 2), see figure 1, and corresponding tiny higher-order peaks could also be indexed with a propagation vector $\mathbf{k} = (0.516, 0, 0)$. Since these peaks have not been observed in the X-ray diffraction patterns they could, in principle, be ascribed to the ordering of La/Ba [10] or they might even be of magnetic origin. This latter hypothesis can be safely excluded, however, since we did not observe any change in the respective peak intensities between 1.5 K and 550 K. As the low-temperature oxygen treatment mainly affects the oxygen stoichiometry, the extra peak could also be ascribed to some kind of oxygen-vacancy and/or charge/stripe ordering phenomenon, at least in the case of the oxygen-defective sample [12,13,14]. For stoichiometric LaBaNiO$_4$ with a nominally single-valent Ni$^{3+}$, this explanation does not seem to be very plausible, however. Another possibility that could explain the very similar intensities of the extra peak in both the as-prepared and the oxidized samples is the existence of a charge-density wave of the type Ni$^{3+\delta}$/Ni$^{3-\delta}$. Such a charge disproportionation, although difficult to observe with diffraction techniques, has been recently reported for other stoichiometric Ni$^{3+}$ compounds close to the boundary between localized and itinerant behaviour, such as AgNiO$_2$ [15] and YNiO$_3$ [16]. In the present LaBaNiO$_{3.85}$ and LaBaNiO$_4$ samples, however, we have not yet been able to establish a model for a charge and/or an oxygen-vacancy ordering that could sufficiently reproduce the observed intensity of the observed extra peak.

It is important to mention that apart from the displacements of the Bragg reflections due to thermal dilatation, the diffraction patterns for fully oxygenated LaBaNiO$_4$ at different temperatures look *exactly the same*. In particular there is no evidence for the appearance of additional reflections at low temperatures that would suggest any type of magnetic order in LaBaNiO$_4$.

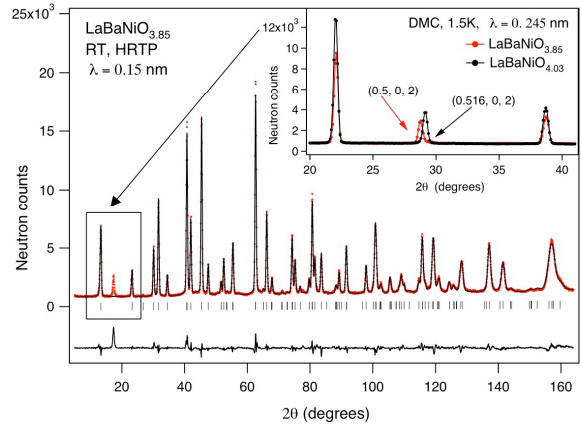

**Figure 1.** Neutron powder-diffraction pattern of LaBaNiO$_{3.85}$ taken at room temperature, together with a Rietveld-refinement fit using the space group I4/mmm. The inset shows the displacement of the (1/2, 0, 2) Bragg reflection to the incommensurate position (0.516, 0, 2) upon increasing the oxygen content to 4.025.

*3.2. Electrical conductivity*

In figure 2 (a) we show the electrical conductivities of the as-prepared samples, plotted as $\sigma$ vs. $T^{-1/4}$ on a semi-logarithmic scale. The measured data for LaSrNiO$_4$, with a slight decrease of $\sigma$ with decreasing temperature, are perfectly in line with corresponding data from the literature [3,8]. As a general trend, the low-temperature conductivity decreases with increasing Ba content, but more important, its temperature dependence changes from a relatively weak one for LaSrNiO$_4$ ($x = 0$) to a $\sigma(T)$ that varies over more than 6 decades from room temperature down to $T = 7$ K for LaBaNiO$_{4-\delta}$ ($x = 1$). In figure 2 (b) we show the corresponding data of the samples that have been annealed at an elevated oxygen pressure. Although there is a change in the data of LaSr$_{0.5}$Ba$_{0.5}$NiO$_{4-\delta}$ and LaBaNiO$_{4-\delta}$ upon oxygen annealing, namely a decrease in the room-temperature conductivity and slight flattening of $\sigma(T)$, the strongly temperature dependent character of the electrical conductivity remains, confirming that stoichiometric LaBaNiO$_4$ is an insulator at $T = 0$. As it was already noticed by Demazeau *et al*. [9] the best physically reasonable fit to the $\sigma(T)$ data of LaBaNiO$_{4-\delta}$ is according to a strongly $T$-dependent 3D variable-range hopping-type (VRH) conductivity, $\sigma(T) = \sigma_0 \exp[-(T_0/T)^{1/4}]$ [17] (see dashed lines in figure 2 (b)), which we will discuss later in section 4.

In the following we will restrict our discussion of the other measured physical properties to the samples LaSrNiO$_4$, the oxygen treated LaSr$_{0.5}$Ba$_{0.5}$NiO$_4$ (as a prototype of an intermediate composition) and LaBaNiO$_4$ to avoid unnecessary complications coming from an off-stoichiometric oxygen content.

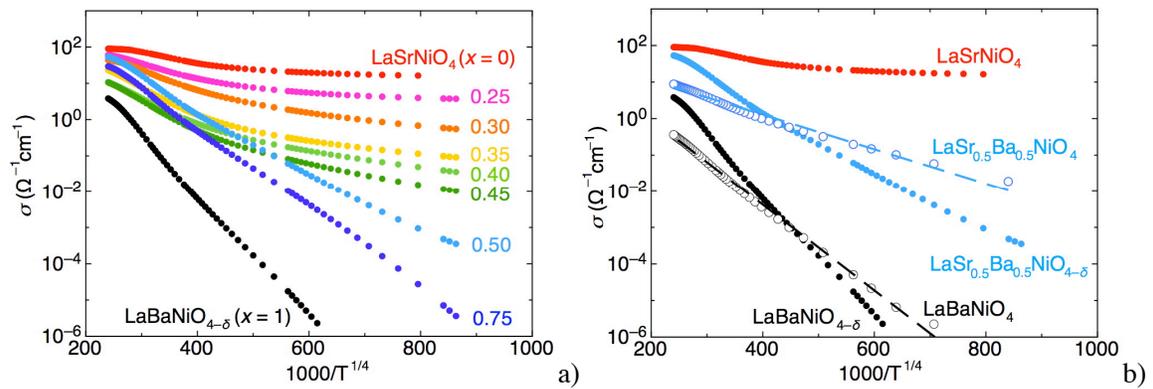

**Figure 2.** (a) Electrical conductivities $\sigma$ vs. $T^{-1/4}$ of as-prepared samples (full circles) and (b) of high-pressure oxygen-annealed samples (open circles). The dashed lines are fits according to a variable-range hopping-type conductivity, see text.

## 3.3. Magnetic susceptibility

In figure 3 we present the magnetic-susceptibility $\chi(T)$ data of LaSrNiO$_4$ and oxygen treated LaSr$_{0.5}$Ba$_{0.5}$NiO$_4$ and LaBaNiO$_4$. Despite the large differences in the respective electrical conductivities, the three curves all look very similar, with an almost constant $\chi_0$ at high temperatures and a small Curie-like upturn in $\chi(T)$ at low temperatures, which is in agreement with the data for LaSrNiO$_4$ from the literature [3]. A corresponding fit to the data above $T = 40$ K assuming $\chi(T) = C/T + \chi_0$ gives an almost universal value $\chi_0 \approx 4 \times 10^{-4}$ emu/mole and a Curie term $C$ of the order of 1–2 $\times 10^{-2}$ emu K/mole (see table 2), which corresponds to only about 5% of the value for free low-spin $S = 1/2$ Ni$^{3+}$ ions. This indicates that the magnetic moments of Ni in electrically insulating LaBaNiO$_4$ are in large majority *not localized*. A mathematically better fit to all the data, including those at temperatures below 40 K, is obtained by allowing for a Néel-type expression $\chi(T) = C/(T+\Theta) + \chi_0$, see figure 3. However, the qualitative results, namely a $\chi_0$ of the order of 3-5 $\times 10^{-4}$ emu/mole and a fairly small Curie term, remain the same. Note that the free-electron value for the Pauli-paramagnetic susceptibility of a metal is $\chi_0 \approx 4 \times 10^{-5}$ emu/mole assuming one charge carrier per formula unit. The observed strong enhancement of the measured $\chi_0$ values over the free-electron value will be discussed below in section 4.

**Table 2.** Fitting results of the $\chi(T)$ data shown in figure 3. The first row for each compound corresponds to a fit according to $\chi(T) = C/T + \chi_0$, the second row to $\chi(T) = C/(T+\Theta) + \chi_0$.

|  | $C$ (emu K/mole) | $\chi_0$(emu/mole) | $\Theta$ (K) |
|---|---|---|---|
| LaSrNiO$_4$ | 1.10(2) $\times 10^{-2}$ | 4.12(2) $\times 10^{-4}$ | - |
|  | 1.62(7) $\times 10^{-2}$ | 3.90(5) $\times 10^{-4}$ | 18.4 ± 1.3 |
| LaSr$_{0.5}$Ba$_{0.5}$NiO$_4$ | 1.60(11) $\times 10^{-2}$ | 4.57(15) $\times 10^{-4}$ | - |
|  | 2.58(30) $\times 10^{-2}$ | 4.23(21) $\times 10^{-4}$ | 25.7 ± 3.2 |
| LaBaNiO$_4$ | 2.41(10) $\times 10^{-2}$ | 4.49(13) $\times 10^{-4}$ | - |
|  | 5.40(60) $\times 10^{-2}$ | 3.41(31) $\times 10^{-4}$ | 43 ± 5 |

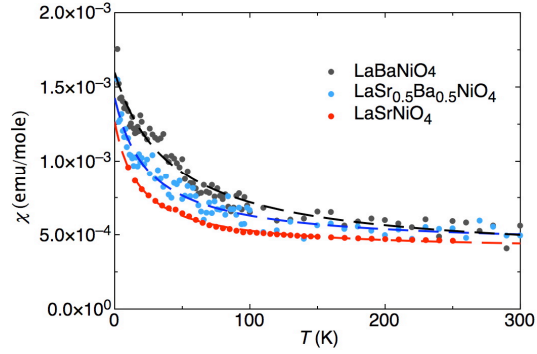

**Figure 3.** Magnetic susceptibilities $\chi(T)$ of LaSrNiO$_4$ (bottom), oxygen treated LaSr$_{0.5}$Ba$_{0.5}$NiO$_4$ (middle) and LaBaNiO$_4$ (upper curve). The dashed lines are fits to the data assuming $\chi(T) = C/(T+\Theta) + \chi_0$ (see text and table 2).

*3.4. Specific heat*

In figure 4 we have plotted the specific-heat data of LaSrNiO$_4$, LaSr$_{0.5}$Ba$_{0.5}$NiO$_4$ and LaBaNiO$_4$ as $C/T$ vs. $T^2$ and for temperatures $T < 16$ K. A standard analysis accounting for a phonon contribution $\beta T^3$ and a linear term $\gamma T$ that is usually ascribed to a finite electron density of states at the Fermi level, gives the parameters presented in table 3. As expected from Debye theory, the Debye temperature $\Theta_D = (233.8 nR/\beta)^{1/3}$ (with $n$ the number of atoms per unit cell and the gas constant $R = 8.314$ J/moleK) decreases for increasing molar mass, i.e., increasing Ba content. While the presence of a linear term in the specific heat of LaSrNiO$_4$ is not unexpected because of its metallic nature, corresponding linear terms of the same order of magnitude are also obtained for LaSr$_{0.5}$Ba$_{0.5}$NiO$_4$ and even for insulating LaBaNiO$_4$, which is rather surprising. The measured Sommerfeld coefficients $\gamma \approx 7\text{-}10$ mJ/moleK$^2$ are all of the same order of magnitude as in metallic LaNiO$_3$ ($\gamma \approx 13$ mJ/moleK$^2$ [18]) but they are large when compared to the free-electron value $\gamma = 2.0$ mJ/moleK$^2$ assuming one charge carrier per formula unit. In section 4 we will relate the measured Sommerfeld coefficients with the corresponding values for the Pauli-paramagnetic susceptibilities $\chi_0$ and the respective free-electron values.

**Table 3.** Fitting results from the specific-heat data in figure 4 according to $C(T) = \beta T^3 + \gamma T$.

|  | LaSrNiO$_4$ | LaSr$_{0.5}$Ba$_{0.5}$NiO$_4$ | LaBaNiO$_4$ |
| --- | --- | --- | --- |
| $\beta$ (mJ/moleK$^4$) | 0.204(1) | 0.273(1) | 0.388(1) |
| $\Theta_D$ (K) | 405 | 368 | 327 |
| $\gamma$ (mJ/moleK$^2$) | 9.15(11) | 7.27(16) | 10.1(1) |

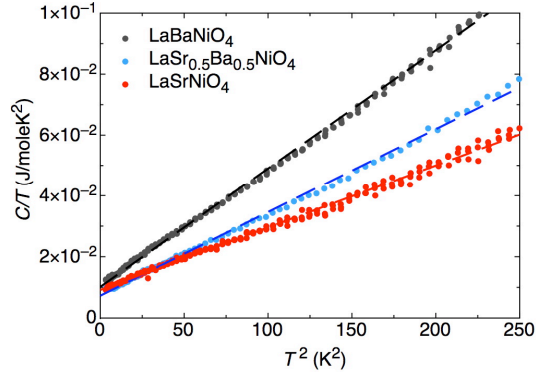

**Figure 4.** Reduced specific-heat $C/T$ vs. $T^2$ data of LaSrNiO$_4$ (bottom), oxygen treated LaSr$_{0.5}$Ba$_{0.5}$NiO$_4$ (middle) and LaBaNiO$_4$ (upper curve). The dashed lines are fits to $C(T) = \beta T^3 + \gamma T$ (see text and table 3).

*3.5. Thermal conductivity*

In section 3.2. we have shown that the electrical transport in LaBaNiO$_4$ is far from being metallic. However, from the magnetic-susceptibility and the specific-heat data alone one cannot distinguish between the physical properties of LaSrNiO$_4$ and those of LaSr$_{0.5}$Ba$_{0.5}$NiO$_4$ and LaBaNiO$_4$, all resembling those of a metal. We therefore carried out additional thermal conductivity measurements on LaSrNiO$_4$ and LaBaNiO$_4$ to probe a further transport property. As can be seen in figure 5, the thermal conductivity of LaSrNiO$_4$ is somewhat larger than that of LaBaNiO$_4$. To be able to make a fair comparison we make use of the kinetic expression for the lattice contribution to the thermal conductivity, $\lambda_{latt} = C_{latt} v l / 2$, where $C_{latt}$ is the lattice specific heat per unit volume, $v$ is the phonon group velocity and $l$ is the phonon mean free path. Since $v$ depends, to first approximation, on the density $\rho$ of a material as $\rho^{-1/2}$, we have to compare the quantities $C_{latt}\rho^{-1/2}l$ to obtain an estimate for the respective lattice contributions. To calculate $C_{latt}$ over the entire temperature range we have used the standard expression for a Debye solid using the above values for $\Theta_D$, and used the X-ray densities $\rho = 6360$ kg/m$^3$ for LaSrNiO$_4$ and $\rho = 6980$ kg/m$^3$ for LaBaNiO$_4$, respectively.

Such a comparison is expected to work best in the high-temperature limit where we may assume that for a given temperature the phonon mean free path $l$ is approximately the same in both isostructural compounds, while $l$ is expected to be limited by extrinsic, sample dependent parameters at low temperatures. A the same time, the Wiedemann-Franz law that links the thermal conductivity to the electrical conductivity is expected to be valid at high temperatures as well. Therefore we focus on the quantity $C_{latt}\rho^{-1/2}$ that should be proportional to $\lambda_{latt}$ at a fixed temperature, e.g. at $T = 290$ K. In figure 5 we have plotted the thus obtained $C_{latt}\rho^{-1/2}$ curves for both compounds in such a way that $C_{latt}\rho^{-1/2}$ for LaBaNiO$_4$ approximately matches the measured thermal-conductivity data between 250 K and 300 K. If we regard the thermal conductivity of this electrically insulating compound as a pure lattice thermal conductivity, the corresponding calculated $C_{latt}\rho^{-1/2}$ curve for LaSrNiO$_4$ should then represent the lattice contribution of the latter compound as well. This calculated phonon contribution to the thermal conductivity is somewhat smaller than the measured data, however (see figure 5). As LaSrNiO$_4$ shows a metallic electrical conductivity, we may make use of the Wiedemann-Franz law to obtain the electronic contribution to its thermal conductivity, $\lambda_{el} = L\sigma T$ with the Lorenz number $L = 2.44 \times 10^{-8}$ WΩK$^{-2}$. Using the room-temperature electrical conductivity $\sigma \approx 90$ Ω$^{-1}$cm$^{-1}$ we obtain $\lambda_{el} \approx 6.6 \times 10^{-2}$ W/Km, which is of the correct order of magnitude to explain the difference between the measured thermal conductivity and the estimated lattice contribution of LaSrNiO$_4$ near room temperature (see figure 5).

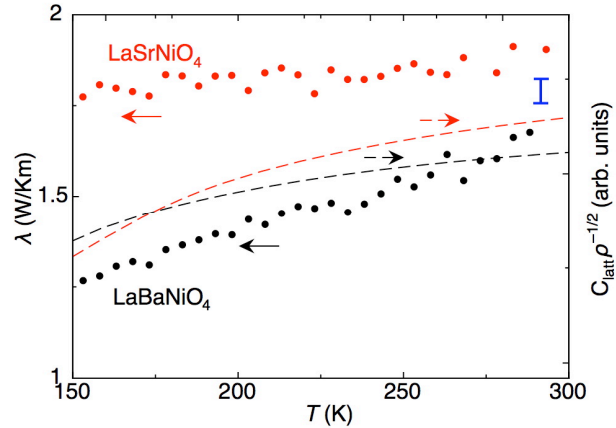

**Figure 5.** Thermal conductivities $\lambda(T)$ of LaSrNiO$_4$ and oxygen treated LaBaNiO$_4$ (left scale). The dashed lines correspond to $C_{latt}\rho^{-1/2}$ that is related to the lattice contributions $\lambda_{latt}$, see text. The solid bar at $T = 290$ K represents the electronic contribution $\lambda_{el}$ for LaSrNiO$_4$ as estimated from the Wiedemann-Franz law near room temperature.

## 4. Discussion

As a main result of the experimental sections we summarize that LaSrNiO$_4$ and LaBaNiO$_4$ represent two end members of the solid solution LaSr$_{1-x}$Ba$_x$NiO$_4$, in which the transport properties gradually change from metallic-like for $x = 0$ to insulating for $x = 1$. By very contrast, the magnetic susceptibility and the specific-heat data clearly resemble those of a metal, with no apparent difference between LaSrNiO$_4$ and LaBaNiO$_4$. We will first discuss the electrical conductivity alone and show that the variable-range hopping (VRH) scenario is certainly valid to explain the corresponding data for LaSr$_{0.5}$Ba$_{0.5}$NiO$_4$ and LaBaNiO$_4$. We then focus on the magnetic susceptibilities and the low temperature specific heats and show that these two quantities are closely related to one another as they are in an ordinary correlated metallic system. We finally briefly discuss a plausible scenario that can explain these seemingly contradictory results.

As we have seen in figure 2 (b) the electrical conductivities of LaSr$_{0.5}$Ba$_{0.5}$NiO$_4$ and in particular that of LaBaNiO$_4$ are very well described by a 3D VRH-type temperature dependence $\sigma(T) = \sigma_0 \exp[-(T_0/T)^\alpha]$ with $\alpha = 1/4$ [17]. Other possible exponents such as $\alpha = 1/3$ (2D VRH [17]), $\alpha = 1/2$ (Coulomb-gap $T$-dependence [19]) or $\alpha = 1$ (activated finite-gap behaviour) result in a considerably worse agreement with the measured data. At first sight, $\sigma(T)$ of the series LaSr$_{1-x}$Ba$_x$NiO$_4$ shown in figure 2 (a) strongly resembles to corresponding data measured on heavily doped semiconductors where the slope $|T_0^{1/4}|$ in a $\ln\sigma$ vs. $T^{-1/4}$ representation progressively decreases as the system approaches the metal-insulator transition with increasing donor concentration (see, e.g., [20]). However, the physics of LaSr$_{1-x}$Ba$_x$NiO$_4$ is expected to be somewhat different from that of such systems where a number of localized states is supposed to be created within a sizeable band gap of an originally insulating or semiconducting system. Preliminary UV excited photoemission experiments that we have done on LaBaNiO$_4$ clearly exclude the presence of such a band gap that would also manifest itself in an activated (i.e., $\ln\sigma \sim -1/T$) temperature dependence of $\sigma(T)$ at high temperatures, which we do not observe. We can use the same argument to exclude a Mott transition upon stretching the crystal lattice by the substitution of Sr by the larger Ba that would also lead to a band gap. As the VRH conduction mechanism, in its original formulation [17], is not restricted to doped semiconductors but more generally assumes the hopping of electrons in a random potential with localized states it may nevertheless be a valid description for the conduction mechanism in LaSr$_{1-x}$Ba$_x$NiO$_4$. In the present context we may mention the examples of Li-substituted and oxygen reduced (La,Sr)$_2$CuO$_4$, respectively [21,22]. While in the former case $T_0 \approx 2 \times 10^6$ K is almost insensitive to the Li-concentration, this quantity increases from $\approx 2000$ K to $\approx 2 \times 10^5$ K as metallic (La,Sr)$_2$CuO$_4$ is progressively reduced. These values are comparable to what we obtain for LaSr$_{0.5}$Ba$_{0.5}$NiO$_4$ and LaBaNiO$_4$ (see table 4).

To check consistency we can relate the parameters $\sigma_0$ and $T_0$ from the VRH expression for $\sigma(T)$ to the electron density of states at the Fermi level $D(E_F)$ as measured by the low-temperature specific heat, to the spatial extension $\xi$ of the quasilocalized wavefunction (localization length), to a mean hopping distance $R$, to the average hopping energy $W$ and to a hopping frequency $\nu$. The quantity $D(E_F)$ can be obtained from the measured Sommerfeld constant using $\gamma = \pi^2 k_B^2 D(E_F)/3$, while $\xi = [18/k_B T_0 D(E_F)]^{1/3}$. The $T$-dependent $R$ becomes $R = [9\xi/8\pi k_B T D(E_F)]^{1/4}$ and $W = 3/4\pi R^3 D(E_F)$. The hopping frequency $\nu = \sigma_0/e^2 R^2 D(E_F)$ should be, in principle, of the order of a typical phonon frequency, i.e., $\approx \nu_D = k_B \Theta_D/h$ (with e the electron charge and $h$ the Planck constant) where we can use our specific-heat results for the Debye temperatures $\Theta_D$. In table 4 we present the corresponding values that we obtain for LaSr$_{0.5}$Ba$_{0.5}$NiO$_4$ and LaBaNiO$_4$.

**Table 4.** Characteristic parameters for the variable-range hopping (VRH) type conductivity in LaSr$_{0.5}$Ba$_{0.5}$NiO$_4$ and LaBaNiO$_4$.

|  | LaSr$_{0.5}$Ba$_{0.5}$NiO$_4$ | | LaBaNiO$_4$ | |
| --- | --- | --- | --- | --- |
|  | 2 K | 300 K | 2 K | 300 K |
| $T_0$ (K) | 1.4 x 10$^4$ | | 5.4 x 10$^5$ | |
| $\sigma_0$ ($\Omega^{-1}$cm$^{-1}$) | 100 | | 210 | |
| $D(E_F)$ (J$^{-1}$m$^{-3}$) | 2.1 x 10$^{47}$ | | 2.8 x 10$^{47}$ | |
| (states eV$^{-1}$/Ni) | 3.1 | | 4.3 | |
| $\xi$ (nm) | 0.76 | | 0.21 | |
| $R$ (nm) | 2.6 | 0.75 | 1.8 | 0.5 |
| $W$ (meV) | 0.40 | 17 | 1.0 | 42 |
| $\nu$ (Hz) | 2.7 x 10$^{11}$ | 3.4 x 10$^{12}$ | 9.6 x 10$^{11}$ | 1.2 x 10$^{13}$ |
| $\nu_D$ (Hz) | 7.7 x 10$^{12}$ | | 6.8 x 10$^{12}$ | |

The values for $\nu$ compare reasonably well with our rough estimate for the hopping frequencies, thereby confirming the validity of our approach. As the above relations are valid for an isotropic 3D system, we have to take the values for $\xi$ and $R$ with the reservation that LaSr$_{1-x}$Ba$_x$NiO$_4$ is certainly anisotropic. The related system (La,Sr)$_2$CuO$_4$ shows an anisotropy of the order of 10-20 of the electronically and magnetically relevant length scales in the metallic regime [23], depending on the crystal direction in which they are measured. The "isotropic" values for $\xi$ may therefore be interpreted in the sense that the true localization lengths along the Ni-O planes are somewhat larger than our calculated $\xi$, i.e., of the order of a few times the lattice constant $a$, while the corresponding smaller localization length in the perpendicular direction is most likely only of the size of the extension of the hybridized Ni-O orbitals in the $c$-direction. The corresponding mean hopping distance is, according to the above estimate, of the same order of magnitude as $\xi$. The average hopping energies are consistent with the requirement $W \sim k_B T$ to make a hopping to distant sites possible.

From the measured Sommerfeld constants $\gamma$ given in table 3 we can calculate the density of states at the Fermi level $D(E_F)$ and compare them with the free-electron gas values $D_{f.e.}(E_F) = (3n/\pi)^{1/3} m_e / \pi \hbar^2$ (with the charge-carrier density $n$, electron mass $m_e$ and $\hbar = h/2\pi$) to obtain the electron effective-mass enhancement $m^*/m_e = D(E_F)/D_{f.e.}(E_F)$. Since we have a nominal $3d^7$ electronic configuration of Ni we may assume here one mobile charge carrier per formula unit to calculate $n = 2/a^2 c$. From $D(E_F)$ and the susceptibility data measured above $T = 40$ K (see table 2) we can also estimate the Stoner enhancement $1/(1-S)$ of the Pauli-paramagnetic susceptibilities using $\chi_0 = 2\mu_B^2 D(E_F)/(1-S)$. In table 5 we show the corresponding values for LaSrNiO$_4$, LaSr$_{0.5}$Ba$_{0.5}$NiO$_4$ and LaBaNiO$_4$.

**Table 5.** Electron density of states, effective mass and Stoner enhancement from specific-heat and magnetic-susceptibility data.

|  | LaSrNiO$_4$ | LaSr$_{0.5}$Ba$_{0.5}$NiO$_4$ | LaBaNiO$_4$ |
|---|---|---|---|
| $D(E_F)$ (J$^{-1}$m$^{-3}$) | 2.66(3) x 10$^{47}$ | 2.07(5) x 10$^{47}$ | 2.81(3) x 10$^{47}$ |
| (states eV$^{-1}$/Ni) | 3.9 | 3.1 | 4.3 |
| $D_{f.e.}(E_F)$ (J$^{-1}$m$^{-3}$) | 5.8 x 10$^{46}$ | 5.7 x 10$^{46}$ | 5.7 x 10$^{46}$ |
| $m^*/m_e$ | 4.62(6) | 3.62(8) | 4.96(5) |
| $1/(1-S)$ | 3.3(1) | 4.6(3) | 3.2(2) |
| $S$ | 0.70(2) | 0.78(5) | 0.69(4) |

These numbers are all comparable to those that have been measured for metallic LaNiO$_3$ ($\gamma \approx 13$ mJ/moleK$^2$, $S \approx 0.58$ [18]) and are usually interpreted as properties of a metal with rather strong electron correlations [4,18]. Our value for $D(E_F) \approx 4$ states per eV and Ni atom for LaSrNiO$_4$ is also in fair agreement with the results from corresponding band-structure calculations for this compound with $D(E_F) \approx 6$ states/eV [4].

The striking similarity of the low-temperature specific heat and the magnetic-susceptibility data of LaSrNiO$_4$, LaSr$_{0.5}$Ba$_{0.5}$NiO$_4$ and LaBaNiO$_4$ as shown in the figures 3 and 4, with almost identical associated material parameters as summarized in table 5, suggests that the underlying physics that determines these parameters does not significantly change in LaSr$_{1-x}$Ba$_x$NiO$_4$ with varying $x$. This implies that LaBaNiO$_4$ has a finite density of states at the

Fermi level, but these states obviously do not contribute to the transport properties, i.e., they must be localized. This situation, together with the observed VRH-type electrical conductivity [17], is very reminiscent of that in a Fermi glass [24,25,26]. In such a system, the Fermi energy $E_F$ lies below a mobility edge $E_c$ that separates localized electronic states from extended states. The free-electron density of states is then replaced by a quasiparticle density of states, but the expressions for the Sommerfeld constant and the Pauli-paramagnetic susceptibility of the system remain formally the same [25]. Very often $E_F$ can be tuned by changing the charge-carrier concentration to transform an insulating Fermi glass ($E_F < E_c$) to a metal ($E_F > E_c$). At this metal-insulator transition (Anderson transition at $E_F = E_c$), the conductivity at $T = 0$ is expected to be a constant $\sigma(0) = Ce^2/\hbar a$ with $C \approx 0.025$–$0.1$ [27], which corresponds to $\sigma(0) \approx 160...640$ $\Omega^{-1}$cm$^{-1}$ in our case. By comparing this estimate with our results shown in figure 2 we conclude that LaSrNiO$_4$ with $\sigma(0) \approx 90$ $\Omega^{-1}$cm$^{-1}$ is indeed at the borderline of an Anderson-like metal-to-insulator transition. This interpretation is supported by measurements of the electrical conductivity on La$_{2-y}$Sr$_y$NiO$_4$ with varying Sr content $y$, where a finite $\sigma(0) \approx 100$ $\Omega^{-1}$cm$^{-1}$ is measured for polycrystalline samples with $y = 1$ [3] and on thin films at the corresponding metal-insulator transition around $y = 0.95$ [8]. With $y$ decreasing from $y \approx 1$, the resistivity $\rho(T) = \sigma(T)^{-1}$ diverges more and more rapidly as $T \to 0$ according to a VRH-type temperature dependence [8], while increasing $y$ above $y \approx 1$ results in a pronounced metallic behaviour in these measurements [3,8]. In LaSr$_{1-x}$Ba$_x$NiO$_4$ we do not expect to significantly change the charge-carrier concentration, however, because we maintain a formal Ni$^{3+}$ oxidation state, and, as a consequence, our experimental values for $D(E_F)$ virtually do not depend on $x$. We therefore conclude that the progressive substitution of Sr by Ba increases $E_c$, rather than changing the charge-carrier concentration or even creating a Mott insulator with a band gap.

In the conventional Anderson metal-to-insulator-transition scenario the magnitude of the mobility edge $E_c$ is determined by the amount of atomic disorder. At first sight it may not be obvious why the isoelectronic substitution of Sr by Ba in LaSr$_{1-x}$Ba$_x$NiO$_4$ that is associated with an expansion of the crystal lattice should lead to a growth of such a disorder, with the stoichiometric LaBaNiO$_4$ as the composition with the largest $E_c$. In a plausible scenario we can relate the atomic disorder to the very large difference between the ionic radii of La and Ba, whereas the respective difference for La and Sr is comparably small. Therefore, atomic disorder is expected to grow with increasing Ba content. However, we want to mention the possibility that charge-localization phenomena that are known to occur in certain other stoichiometric Ni-oxides [15,16] may also play a certain role in LaSr$_{1-x}$Ba$_x$NiO$_4$. In this spirit we can tentatively relate the observed superstructure peaks in our oxygen stoichiometric LaBaNiO$_4$ to such a charge-ordering phenomenon. Whether charge localization alone can lead to a Fermi-glass type behaviour or not is not known to us. However, such a scenario could well explain that the charge-ordered $La_{2-y}Sr_yNiO_4$ samples for $y \leq 0.7$ (see, e.g., [28] and the cited references therein) show a VRH-type electrical conductivity [8, 29], but also a linear term in their low-temperature specific heats ($\gamma \approx 10$ mJ/moleK$^2$) and Pauli-like paramagnetic susceptibilities at high temperatures that are of the order of $\chi_0 \approx 6 \times 10^{-4}$ emu/mole [29]. These observations, that are very similar to ours, may suggest that an Anderson-like transition is realized in $La_{2-y}Sr_yNiO_4$ around $y \approx 1$, mainly (but perhaps not exclusively) by tuning the Fermi energy with varying $y$, while in $LaSr_{1-x}Ba_xNiO_4$ it is most likely predominantly the mobility edge that is being tuned.

We conclude that we observe an Anderson-type metal-to-insulator transition in $LaSr_{1-x}Ba_xNiO_4$ around $x \approx 0$ that leads to a Fermi glass behaviour of $LaBaNiO_4$, with insulating transport properties at $T = 0$ but at the same time with a finite density of states at the Fermi level. We suggest that this transition is caused by an increase of the mobility edge in conjunction with the progressive substitution of Sr by Ba and the increasing atomic disorder associated with it. The insulating state in $LaBaNiO_4$ does neither show antiferromagnetic order nor a finite band gap. We expect that a doping with charge carriers on the moderate level that is sufficient to induce metallic behaviour in the copper oxides (i.e., $\approx 0.15$ holes per transition-metal atom [1]) will not lift the localization of the electronic states in the Fermi-glass state and will therefore not lead to metallic transport properties.

**Acknowledgements**

We would like to thank to S. Siegrist, B. Bischof, L. Keller, K. Conder and V. Pomjakushin for their technical assistance and to T. Brugger for providing the photoemission data on $LaBaNiO_4$. This work was supported in part by the Swiss National Foundation through the NCCR MaNEP and Grant. No. 20-111653.

**References**

[1] Bednorz J G and Müller K A, 1986 *Z. Physik* B **64** 189

[2] Jin S, Tiefel T H, McCormack M, Fastnacht R A, Ramesh R and Chen L H, 1994 *Science* **264** 413

[3] Cava R J, Batlogg B, Palstra T T, Krajewski J J, Peck Jr. W F, Ramirez A P and Rupp L W, 1991 *Phys. Rev.* B **43** 1229

[4] Anisimov V I, Bukhvalov D and Rice T M, 1999 *Phys. Rev.* B **59** 7901

[5] Kuiper P, Kruizinga G, Ghijsen J, Sawatzky G A and Verweij H, 1989 *Phys. Rev. Lett.* **62** 221

[6] van Elp J, Eskes H, Kuiper P and Sawatzky G A, 1992 *Phys. Rev.* B **45** 1612

[7] Eisaki H, Uchida S, Mizokawa T, Namatame H, Fujimori A, van Elp J, Kuiper P, Sawatzky G A, Hosoya S and Katayama-Yoshida H, 1992, *Phys. Rev.* B **45** 12513

[8] Shinomori S, Okimoto Y, Kawasaki M and Tokura Y, 2002 *J. Phys. Soc. Jpn.* **71** 705

[9] Demazeau G, Marty J L, Buffat B, Dance J M, Pouchard M, Dordor P and Chevalier B, 1982 *Mat. Res. Bull.* **17** 37

[10] Alonso J A, Amador J, Gutiérrez-Puebla E, Monge M A, Rasines I, Ruiz-Valero C and Campa J A, 1990 *Solid State Commun.* **12** 1327

[11] Demazeau G, Pouchard M and Hagenmüller P, 1976 *J. Solid State Chem.* **18** 159

[12] Tranquada J M, Axe J D, Ichikawa N, Nakamura Y, Uchida S and Nachumi B, 1996 *Phys. Rev.* B **54** 7489

[13] Chen C H, Cheong S W and Cooper A S, 1993 *Phys. Rev. Lett.* **71** 2461

[14] Medarde M and Rodriguez-Carvajal J, 1997 *Z. Phys.* B **102** 307

[15] Wawrzynska E, Coldea R, Wheeler E M, Mazin I I, Johannes M D, Sörgel T, Jansen M, Ibberson R M and Radaelli, 2007 *Phys. Rev. Lett.* **99** 157204

[16] Alonso J A, García-Muñoz J L, Fernández-Díaz M T, Aranda M A G, Martínez-Lope M J and Casais M T, 1999 *Phys. Rev. Lett.* **82** 3871

[17] Mott N F, 1968 J. Non-Cryst- Solids **1** 1

Mott N F, 1969 *Phil. Mag.* **19** 835

[18] Sreedhar K, Honig J M, Darwin M, McElfresh M, Shand P M, Xu J, Crooker B C and Spalek J, 1992 *Phys. Rev.* B **46** 6382

[19] Shklovskii B I and Efros A L, 1984 *Electronic properties of doped semiconductors* (Berlin: Springer).

[20] Shafarman W N and Castner T G, 1986 *Phys. Rev.* B **33** 3570

[21] Kastner M A, Birgeneau R J, Chen C Y, Chiang Y M, Gabbe D R, Jenssen H P, Junk T, Peters C J, Picone P J, Tineke T, Thurston T R and Tuller H L, 1988 *Phys. Rev.* B **37** 111

[22] Osquiguil E J, Civale L, Decca R and de la Cruz F, 1988 *Phys. Rev.* B **38** 2840

[23] Kohout S, Schneider T, Roos J, Keller H, Sasagawa T and Takagi H, 2007 *Phys. Rev.* B **76** 064513

[24] Anderson P W, 1970 *Comments on Solid State Physics* **2** 193

[25] Freedman R and Hertz J A, 1977 *Phys. Rev.* B **15** 2384

[26] Müller K A, Penney T, Shafer M W and Fitzpatrick W J, 1981 *Phys. Rev. Lett.* **47** 138

[27] Mott N F, 1972 *Phil. Mag.* **26** 1015

[28] Ishizaka K, Arima T, Murakami Y, Kajimoto R, Yoshizawa H, Nagaosa N and Tokura Y, 2004 *Phys. Rev. Lett.* **92** 196404

[29] Kato M, Maeno Y and Fujita T, 1991 *J. Phys. Soc. Japan* **60** 1994